\documentclass[prd,aps,nofootinbib,showpacs, 10 pt]{revtex4}
\usepackage{amsmath}
\usepackage{amsmath,graphicx,color,epsfig}

\usepackage{ifthen}
\usepackage{epsfig}
\usepackage{amssymb}
\usepackage{graphicx}
\usepackage{amsmath}
\usepackage[center]{subfigure}

\makeatletter
\def\fmslash{\@ifnextchar[{\fmsl@sh}{\fmsl@sh[0mu]}}
\def\fmsl@sh[#1]#2{%
  \mathchoice
    {\@fmsl@sh\displaystyle{#1}{#2}}%
    {\@fmsl@sh\textstyle{#1}{#2}}%
    {\@fmsl@sh\scriptstyle{#1}{#2}}%
    {\@fmsl@sh\scriptscriptstyle{#1}{#2}}}
\def\@fmsl@sh#1#2#3{\m@th\ooalign{$\hfil#1\mkern#2/\hfil$\crcr$#1#3$}}
\makeatother

\begin{document}
\title{Charmed Scalar Meson Production in $B$ Decays}
\author{Yue-Long Shen$^a$ and Xin Yu$^b$}

\affiliation{\it $^a$ College of Information Science and
Engineering,
Ocean University of China, Qingdao, Shandong 266100, P.R. China\\
$^b$ Institute of High Energy Physics, CAS, P.O. Box 918(4),
100049, P.R. China}

\begin{abstract}

The study on the  charmed scalar meson spectroscopy has become a
hot topic both experimentally and theoretically. The $B_{(s)}$
decays provide an ideal place to study their property. We employ
the $B$-meson light-cone sum rules to compute the $\bar B_s^0\to
D_s^{*+}(2317)$ and  $B^-\to D_0^{*0}(2400)$ transition form
factors at large recoil, assuming $D_s^{*+}(2317)$ and
$D_0^{*0}(2400)$ being scalar quark-anti-quark states. The results
are extrapolated  to the whole momentum region with the help of
HQET.  Considering large uncertainties, our results can be
consistent with the previous studies, while  the power corrections
should be large. We also estimate the semi-leptonic decays $\bar
B_s^0\to D_s^{*+}(2317)l \bar{\nu}_l$ and  $B^-\to D_0^{*0}(2400)l
\bar{\nu}_l$.  The branching fraction of the semi-leptonic $\bar
B_s^0\to D_s^{*+}(2317)l \bar{\nu}_l$ decay is around  $6\times
10^{-3} $ for light leptons and   $0.8\times 10^{-3} $ for tau
final state.  The predicted branching ration of $B^-\to
D_0^{*0}(2400)l \bar{\nu}_l$ is slightly larger than $\bar
B_s^0\to D_s^{*+}(2317)l \bar{\nu}_l$, and we hope the future data
in LHCb can test these results.

\end{abstract}

\maketitle

\section{Introduction}

The charmed scalar meson spectroscopy has evoked many interests
since the observation of $D_{s0}^*(2317)$ by Babar collaboration
at 2003. In addition, the signal for the isospin doublet
$D_{0}^*(2400)$ has also been reported by
Belle\cite{Abe:2003zm}and Focus\cite{Link:2003bd}in the $D\pi$
final state. Recently more measurements on the charmed scalar
meson final state in $B$ decays have been
performed\cite{Aubert:2009wg}. The low
 mass and the narrow width of $D_{s0}^*(2317)$ indicates some
 hints  on its mysterious inner structure. It is regarded as a
 scalar meson state in some studies, while it has also been
 assigned to be a four-quark state or the molecular state.
 Until now, the structure of $D_{s0}^*(2317)$ is
 still a controversial problem. As for
 $D_{0}^*(2400)$, there is less information from experiments,
  and our knowledge of its property is even poorer. So we need more
  phenomenological analysis to clarify the inner structure of these $p$-wave states.

A great number of $B$ decay events have been  accumulated at $B$
factories which provide  good places to test the inner  structure
of the charmed scalar meson. To study the $B$-to-scalar meson
decay modes theoretically, an essential task is to evaluate $ \bar
B_s^0\to D_s^+(2317)$ and $ B^-\to D_0^{*0}(2400)$ transition form
factors. In heavy quark effective theory(HQET)\cite{Isgur}, the
heavy-to-heavy form factor can be reduced to the universal
Isgur-Wise(IW) function $\xi (v\cdot v')$ in the heavy quark
limit. In order to estimate the form factors or the IW function,
one must employ the non-perturbative methods. There have existed
some phenomenological studies using different approaches,
including the phenomenological model \cite{Zhao:2006at}, the QCD
sum rules approach \cite{Huang:2004et,Aliev:2006qy,Azizi:2008tt},
PQCD approach \cite{Li:2008ts}, Lattice QCD
\cite{Hashimoto:1999yp,de Divitiis:2007ui,de Divitiis:2007uk}, as
well as the light-cone sum rules (LCSR)\cite{Li:2009wq}.

LCSR \cite{LCSR 1, LCSR 2,LCSR 3}combines the traditional QCD sum
rules \cite{SVZ} with the theory of hard exclusive process, and
offers a systematic way to compute the soft contribution to the
transition form factor. The vacuum-to-hadron correlation function
is computed in terms of light-cone OPE in the LCSR. In the
conventional LCSR for $ \bar B_s^0\to D_s^+(2317)$ form factor,
the correlation function is taken between the vacuum and
$D_s^+(2317)$ state, whereas the $B$ meson is interpolated by a
local current. The long distance effect of the form factor is then
described by the distribution amplitudes(DAs) of $D_s^+(2317)$. As
the structure of $D_s^+(2317)$ is not well understood, the DAs of
$D_s^+(2317)$ are rather model dependent. In this paper, we employ
a different sum rule for the transition form factor following
Ref.\cite{KMO}, where the correlation function is constructed with
the on-shell $B$-meson and the interpolated current for the
charmed scalar meson. As the nonperturative dynamics is
parameterized in terms of the $B$-meson DAs\cite{GN,Grozin}, the
new method is  usually called $B$-meson LCSR and it has been
widely applied to the calculation of heavy-to-light matrix
elements\cite{Wang:2009hra,Khodjamirian:2010vf}.

In this work, we will employ the $B$-meson LCSR approach to
evaluate the the $\bar B_s^0\to D_{s0}^{*+}(2317)$ and $ B^-\to
D_0^{*0}(2400)$ form factors. In our calculation
$D_{s0}^{*+}(2317)$ and $D_0^{*0}(2400)$ are regarded as $q \bar
q$ mesonic states. The relevant semi-leptonic $\bar B_s^0\to
D_{s0}^{*+}(2317)l\nu$ and $ B^-\to D_0^{*0}(2400)l\nu$ decay
modes are also  analyzed. The large number of data accumulated in
the $B$ factories and LHC-b can test whether our assumption is
reasonable, and the result can help to clarify the inner
structures of the new measured charmed scalar mesons.

The  paper is arranged as follows: We firstly derive the LCSR for
the $\bar B_s^0\to D_{s0}^{*+}(2317)$ and $B^-\to D_0^{*0}(2400)$
form factors in the section II. The contributions from both
two-particle and three-particle wave functions of $B$ meson are
computed. The numerical analysis of LCSR for the transition form
factors at large recoil region is displayed in section III. The
HQET is adopted to describe  transitions at the small recoil
region. Moreover, detailed comparisons between  the form factors
obtained under various approaches are also presented here.
Utilizing these form factors, the  branching fractions of
semileptonic decays are calculated in section IV.  The last
section is devoted to the conclusion.

\section{the light-cone sum rules for form factors}

The $B$-to-charmed scalar meson transition form factor induced by
an axial vector current is defined by:
\begin{eqnarray}
&&\langle D^*_0(p)|\bar{c}\gamma_\mu\gamma_5 b|
\bar{B}(p+q)\rangle =-i \left \{p_\mu f^+_{BD^*_0}(q^2)+q_\mu
f^-_{BD^*_0}(q^2) \right \},\label{formdef}
\end{eqnarray}
where the notation ``$\bar{B}$" denotes $\bar{B}^0$,$B^+$ and
$\bar{B}_s$, and $D^*_0$ refers to $D_{s0}^{*+}(2317)$ and
$D_0^{*0}(2400)$. To obtain the form factors with $B$ meson LCSR,
we consider the following correlation function with   on-shell
$B$-meson state:
\begin{equation}
F_{\mu}(p,q)= i\int d^4x ~e^{i p\cdot x} \langle
0|T\left\{\bar{q}(x) c(x), \bar{c}(0)\gamma_\mu(1-\gamma_5)
b(0)\right\}|\bar{B}(P+q)\rangle\,, \label{eq-corr}
\end{equation}
where $\bar{c}\gamma_\mu(1-\gamma_5) b$ is the $b \to c$
(electro)weak currents and $\bar{q}c$ is the interpolating current
for a charmed scalar meson.

The hadronic representation of the correlation function can be
written as
\begin{eqnarray}
F_{\mu}(p,q)&=&\frac{\langle0|\bar q(0)c(0)| D_{0}^{*}(p)\rangle
\langle D_{0}^{*}(p)|\bar c(0)\gamma_{\mu}\gamma_5
b(0)|\bar B(P+q)\rangle}{m^2_{D_{0}^{*}}-p^2}\nonumber\\
&&+\sum_h \frac{\langle0|\bar q(0)c(0)| h(p)\rangle \langle
h(p)|\bar c(0)\gamma_{\mu}\gamma_5 b(0)|\bar
B(P+q)\rangle}{s-p^2}. \label{eq:cofunh}
\end{eqnarray}
 The decay constants
$f_{D_{0}^{\ast}}$ and $\widetilde{f}_{D_{0}^{\ast}}$
 are given by
 \begin{eqnarray}
 \langle 0|\bar q \gamma_{\mu}
 c|D^{\ast}_{0}(p)\rangle=f_{D_{0}^{\ast}}p_{\mu}\;,\;
 \langle 0|\bar q
 c|D^{*}_{0}(p)\rangle=m_{D^{*}_{0}}\widetilde{f}_{D_{0}^{\ast}},
\end{eqnarray}
where $f_{D^*_{0}}=(m_c-m_q)\widetilde{f}_{D^*_{0}}/m_{D^*_{0}}$
and $m_c$ , $m_q$ are the current masses of charm quark and light
quark, respectively. Inserting the definitions of the form factors
and decay constants, the correlation function reads:
\begin{eqnarray}
 F_{\mu}(p,q)&=&\frac{-im^2_{D_{0}^{*}}f_{D_{0}^{*}}}{(m_c-m_q)(m^2_{D_{0}^{*}}-p^2)}
 [f_{D^*_{s0}}^+(q^2)p_{\mu}+f_{D^*_{0}}^-(q^2)q_{\mu}]
 \nonumber \\
 &&+\int_{s_0^{ D^*_{0}}}^{\infty}ds
 \frac{\rho_+^h(s,q^2)p_{\mu}+\rho_-^h(s,q^2)q_{\mu}}{s-p^2},
\end{eqnarray}
where  $s_0^{D^*_{0}}$ is the threshold parameter corresponding to
the $D^*_{0}$ channel.

On the other side, in the deep Euclidean region, the correlation
function can be calculated in the perturbative theory using the
operator production expansion near the light cone :
\begin{eqnarray}
 F_{\mu}(p,q)&=&F_+^{\rm{QCD}}(q^2,p^2)p_{\mu}+F_-^{\rm{QCD}}(q^2,p^2)q_{\mu} \\
 &=&\int_{m_c^2}^{\infty}ds\frac{1}{\pi}\frac{\rm{Im}F_+^{\rm{QCD}}(q^2,p^2)}{s-p^2}p_{\mu}
 +\int_{m_c^2}^{\infty}ds\frac{1}{\pi}\frac{\rm{Im}F_-^{\rm{QCD}}(q^2,p^2)}{s-p^2}q_{\mu}. \nonumber \label{eq:Piquark}
\end{eqnarray}
Applying the  quark-hadron duality
\begin{eqnarray}
 \rho_i^h(s,q^2)=\frac{1}{\pi}{\rm{Im}}F_i^{\rm{QCD}}(q^2,p^2)\Theta(s-s_0^h),\label{eq:rhospec}
\end{eqnarray}
with $i=``+,-"$ and  performing Borel transformation with respect
to the variable $p^2$,  we can derive the sum rules for the form
factors  as
\begin{eqnarray}
 f_i(q^2)=-i\frac{m_c-m_q}{\pi
 f_{D^*_{0}}m^2_{D^*_{0}}}\int_{m_c^2}^{s_0^h}ds\;{\rm{Im}}F_i^{\rm{QCD}}(q^2,s){\rm exp}\bigg(\frac{m^2_{D^*_{0}}-s}{M_B^2}\bigg).
 \label{sum rules for Ds(2317)}
\end{eqnarray}
The leading-order contribution to the OPE is illustrated in
Fig.~1a. The correlation function can be calculated by contracting
the charm quark fields in Eq. (\ref{eq-corr}) and inserting the
$c$ quark propagator, then we arrive at:
\begin{eqnarray} F_{\mu}^{(B)}(p)&=& i\int d^4x ~e^{i p\cdot
x}\int\frac{d^4k}{(2\pi)^4}{ie^{-ik\cdot x}} \langle
0|T\left\{\bar{q}(x) S_F(x,0) \gamma_\nu(1-\gamma_5)
b(0)\right\}|\bar{B}(P_B)\rangle\,\label{creq}
\end{eqnarray}


\begin{figure}[t]
\begin{center}
\includegraphics[width=8cm]{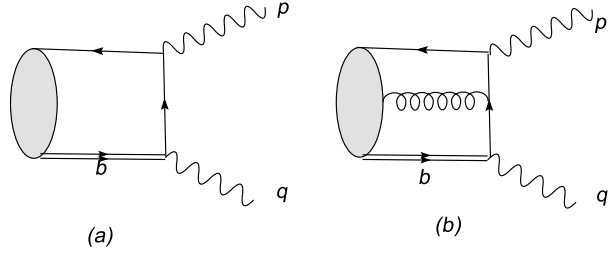}\\
\end{center}
\caption{ \it Diagrams corresponding to the contributions of (a)
two-particle and (b) three-particle $B$ -meson DA's to the
correlation function (\ref{eq-corr})} \label{fig-diags}
\end{figure}


The full quark propagator can be written
as\cite{cite:background1},
\begin{eqnarray}
S_F(x,0)_{ij}&=& \delta_{ij}\int {d^4 k \over (2 \pi)^4} e^{-i
kx}{i \over \not \! k -m_c} -i g \int {d^4 k \over (2 \pi)^4}
e^{-i kx} \int_0^1 d\alpha [{1 \over 2} {\not\! k
+m_c \over (m_c^2 -k^2)^2} G^{\mu \nu}_{ij}(\alpha x)\sigma_{\mu \nu }\nonumber \\
&& +{1 \over m_c^2-k^2}\alpha x_{\mu} G^{\mu \nu}(\alpha
x)\gamma_{\nu}],
\end{eqnarray}
where the first term is the free-quark propagator and $G^{\mu
\nu}_{i j}=G_{\mu \nu}^{a} T^a_{ij}$ with ${\mbox{Tr}}[T^a T^b]={1
\over 2}\delta^{ab}$. Inserting this propagator to
Eq.(\ref{creq}), we can find that the long distance contribution
to the correlation function is expressed by non-local matrix
elements, which defines the $B$-meson light-cone DA. In the
leading Fock state:
\begin{eqnarray}
&& \langle 0|\bar{q}_{2\alpha}(x)[x,0] h_{v\beta}(0)
|\bar{B}_v\rangle
\nonumber \\
&& = -\frac{if_B m_B}{4}\int\limits _0^\infty d\omega e^{-i\omega
v\cdot x} \left [(1 +\not\!v) \left \{ \phi^B_+(\omega) -
\frac{\phi_+^B(\omega) -\phi_-^B(\omega)}{2 v\cdot x} \not\! x
\right \}\gamma_5\right]_{\beta\alpha} \,, \label{eq-BDAdef}
\end{eqnarray}
where $[x,0]$ is the  path-ordered gauge factor. The variable
$\omega>0$ is the plus component of the spectator-quark momentum
in the $B$ meson. The three-particle DAs' contribution is shown in
the  diagram  Fig.~(1b), with the definition
\begin{eqnarray}
&&\langle 0|\bar{q_2}_\alpha(x) G_{\lambda\rho}(ux)
h_{v\beta}(0)|\bar{B}^0(v)\rangle=
\frac{f_Bm_B}{4}\int\limits_0^\infty d\omega \int\limits_0^\infty
d\xi\,  e^{-i(\omega+u\xi) v\cdot x}
\nonumber \\
&&\times \Bigg [(1 +\not\!v) \Bigg \{
(v_\lambda\gamma_\rho-v_\rho\gamma_\lambda)
\Big(\Psi_A(\omega,\xi)-\Psi_V(\omega,\xi)\Big)
-i\sigma_{\lambda\rho}\Psi_V(\omega,\xi)
\nonumber\\
&&-\left(\frac{x_\lambda v_\rho-x_\rho v_\lambda}{v\cdot
x}\right)X_A(\omega,\xi) +\left(\frac{x_\lambda \gamma_\rho-x_\rho
\gamma_\lambda}{v\cdot
x}\right)Y_A(\omega,\xi)\Bigg\}\gamma_5\Bigg]_{\beta\alpha}\,,
\label{eq-B3DAdef}
\end{eqnarray}
where the gauge link factors are omitted for brevity. The DA's
$\Psi_{V}$,$\Psi_{A}$, $X_A$ and $Y_A$ depend
 on two variables $\omega$ and $\xi$, corresponding to the plus components of the light-quark and gluon momenta
in the $B$ meson.

Substituting the $B$ meson distribution function into the
correlation function and employing  the quark hadron
duality(\ref{eq:rhospec}), we arrive at the sum rules for
transition form factors as
\begin{eqnarray}
f_{BD^*_0}^+&=&\frac{f_{B}m_{B}(m_c-m_s)}{f_{D^*_{0}}m^2_{D^*_{0}}}\int_0^{\sigma_0}d\sigma
e^{-(s-m_{D_{0}^{*2}})/M^2}\{m_B(\bar{\sigma}-r_c)(
\frac{1}{\bar{\sigma}}+\frac{m_Bm_c}{\bar{\sigma}^2m_B^2+m_c^2-q^2})\phi_+-\frac{m_B^2m_c(\bar{\sigma}-r_c)}{\bar{\sigma}^2m_B^2+m_c^2-q^2}\phi_-\nonumber
\\&&+[-\frac{1}{\bar{\sigma}}-\frac{m_Bm_c}{\bar{\sigma}^2m_B^2+m_c^2-q^2}+\frac{2m_B^3m_c\bar{\sigma}(\bar{\sigma}-r_c)}
{(\bar{\sigma}^2m_B^2+m_c^2-q^2)^2}]\Phi_\pm\}+f_{BD^*_0}^{+3p},\end{eqnarray}
\begin{eqnarray}
f_{BD^*_0}^-&=&-\frac{f_{B}m_{B}(m_c-m_s)}{f_{D^*_{0}}m^2_{D^*_{0}}}\int_0^{\sigma_0}d\sigma
e^{-(s-m_{D_{0}^{*2}})/M^2}\{m_B(\sigma+r_c)(
\frac{1}{\bar{\sigma}}+\frac{m_Bm_c}{\bar{\sigma}^2m_B^2+m_c^2-q^2})\phi_+-\frac{m_B^2m_c(\sigma+r_c)}{\bar{\sigma}^2m_B^2+m_c^2-q^2}\phi_-\nonumber
\\&&+[\frac{1}{\bar{\sigma}}+\frac{m_Bm_c}{\bar{\sigma}^2m_B^2+m_c^2-q^2}+\frac{2m_B^3m_c\bar{\sigma}(\sigma+r_c)}
{(\bar{\sigma}^2m_B^2+m_c^2-q^2)^2}]\Phi_\pm\}+f_{BD^*_0}^{-3p},\label{eq:formre2}
\end{eqnarray}
where the argument of the wave functions is $m_B\sigma$. In
addition, $\bar{\sigma}=1-\sigma$ and $\sigma_0$ is the root of
the equation
$\bar{\sigma}s_0-(\sigma\bar{\sigma}+r_c^2)m_{B}^2+\sigma
m_{B}^2q^2=0$. The modified wave function
$\Phi_\pm(\omega)=\int^\omega_0d\tau[\phi_+(\tau)-\phi_-(\tau)]$.
The contributions from  three particle $B$ meson DAs are denoted
by $f_{BD^*_0}^{+3p}$ and $f_{BD^*_0}^{-3p}$, which are given in
the appendix.
\section{Numerical analysis of sum rules for form factors}

Now we are going to calculate the form factors $f_{D^*_0}(q^2)$
and $f_{D^*_{s0}}(q^2)$ numerically.  In the following, we list
the relevant input parameters for the $D_{s}^+(2317)$ and
$D_{0}^*(2400)$. Their mass is taken from PDG \cite{pdg}:
$m_{D^*_{s0}}=2.318{\rm GeV}$ and $m_{D^*_{0}}=2.318{\rm GeV}$.
The decay constant $\tilde{f}_{D^*_{s0}}=(250 \pm 25 ){\rm
MeV}$\cite{Colangelo:2005hv}. For the $D_{0}^*(2400)$ state, we
expect $f^i_{D_0}/f^i_{D_{s0}}=f^i_{D}/f^i_{D_{s}}$ in the SU(3)
limit. We adopt the values $f_D=(223\pm 18){\rm MeV}$ and
$f_{D_s}=(274\pm 20){\rm MeV}$, we find
$\tilde{f}_{D^*_{0}}=(203\pm 30){\rm MeV}$.
As for the decay constant of $B_s$ meson,  we use the results
$f_B=130 {\rm MeV}$ \cite{Aliev:1983ra} and $f_{B_s}/f_{B} = 1.16
\pm 0.09$ \cite{Colangelo:2000dp} determined from QCDSR. The
 threshold parameter $s_0$ can be fixed by fitting the LCSR of the
 charmed meson masses to  the experimental data.
Numerically,   the threshold value in the $X$ channel would be
$s^0_{X}=(m_X+\Delta_X)^2$, where $\Delta_X$ is about $0.6$ GeV
\cite{dosch,Wang:2007ys, Colangelo}, and we simply take it as
$(0.6 \pm 0.1)\; \mathrm{GeV}$ in the error analysis. The
two-particle DAs of $B$-meson inspired from QCD sum rule analysis
reads \cite{GN}:
\begin{eqnarray}
\phi_+^B(\omega) & = & \dfrac{\omega}{\omega_0^2}\,e^{-\frac{\omega}{\omega_0}}\,,\nonumber\\
\phi_-^B(\omega) & = &
\dfrac{1}{\omega_0}\,e^{-\frac{\omega}{\omega_0}}\,, \label{eq-GN}
\end{eqnarray}
and the 3-particle DAs are given by:
\begin{eqnarray}
\Psi_A(\omega,\,\xi)& =& \Psi_V(\omega,\,\xi) \,=\,
\dfrac{\lambda_E^2 }{6\omega_0^4}\,\xi^2
e^{-(\omega\,+\,\xi)/\omega_0}\,,
\nonumber\\
X_A(\omega,\,\xi)& = & \dfrac{\lambda_E^2 }{6\omega_0^4}\,
\xi(2\omega-\xi)\,e^{-(\omega\,+\,\xi)/\omega_0}\,,\nonumber\\
Y_A(\omega,\,\xi)& =&  -\dfrac{\lambda_E^2 }{24\omega_0^4}\,
\xi(7\omega_0-13\omega+3\xi)e^{-(\omega\,+\,\xi)/\omega_0}\,.
\label{eq-3partexp}
\end{eqnarray}
The parameters $\omega_0$, $\lambda_H$ and $\lambda_E$  satisfy
the conditions adopted in\cite{GN}:
\begin{eqnarray}
\omega_0=\frac23 \bar{\Lambda},
~~\lambda_E^2=\lambda_H^2=\frac32\omega_0^2= \frac23
\bar{\Lambda}^2\,. \label{eq-paramrel} \end{eqnarray} Numerically
we employ the values $\omega^B_0=0.45\pm0.10$GeV and
$\omega^{B_s}_0=0.50\pm0.10$GeV,  here we have taken small SU(3)
breaking effect into account.

After fixing the corresponding parameters,  we can proceed to
compute the numerical values of the form factors. In principle,
the form factors should not depend on the the unphysical Borel
mass $M^2$. However, the OPE series are truncated up to next to
leading Fock state of the $B$ meson and the QCD corrections are
not considered, a manifest dependence of the form factors on the
Borel parameter $M^2$ would emerge. Therefore, we should search
for the so-called ``Borel window", where  Borel mass dependence is
mild, in order that the truncation is acceptable.

\begin{figure}
\begin{center}
\includegraphics[width=18.cm]{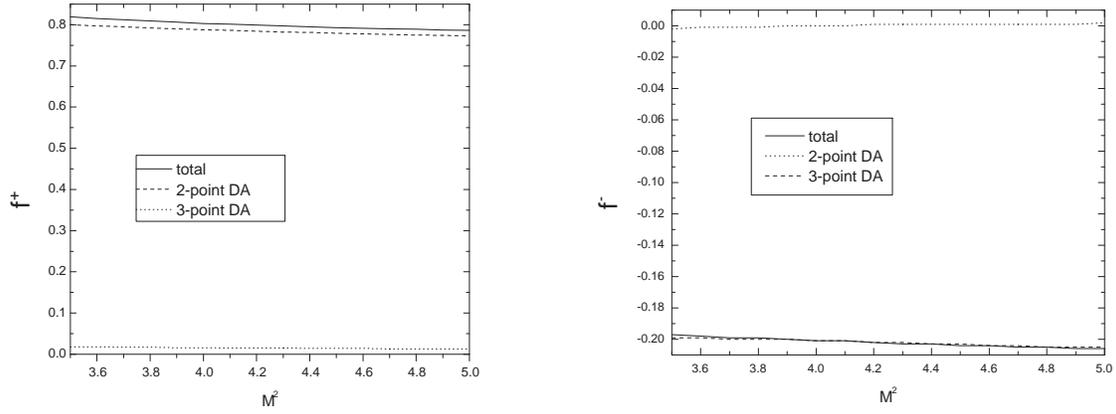}
\vspace{-1.5cm} \caption{ The dependence form factor
$f_{D^*_{s0}}^{+}(0)$  and $f_{D^*_{s0}}^-(0)$  on the Borel mass
$M^2$, the contribution from 2-point $B$ meson DA is denoted by
the dashed line, and the dotted line represents the 3-point DA
contribution. The solid line gives the total results.}
 \label{fig:mtdependence}
 \end{center}
 \end{figure}

\begin{figure}
\begin{center}
\includegraphics[width=18.cm]{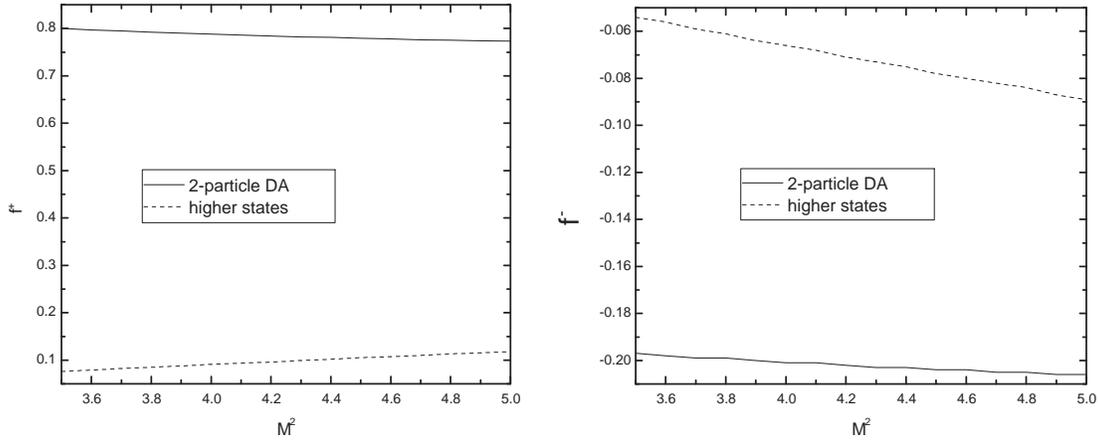}
\vspace{-1.5cm} \caption{ The dependence form factor
$f_{D^*_{s0}}^{+}(0)$  and $f_{D^*_{s0}}^-(0)$ on the Borel mass
$M^2$, the contribution of higher exited states and the continuum
states in the whole sum rules is shown by the dotted line.}
 \label{fig:higherstates}
 \end{center}
 \end{figure}

We firstly focus on the form factors at zero momentum transfer. To
extract the form factor $f_{D_0^*}^i(0)$,   the contribution from
the higher resonances and continuum states should be less than 30
\% in the total sum rules and the value of $f_{D_0^*}^i(0)$ should
not be sensitive to the Borel mass. In view of these
considerations, the Borel parameter $M^2$ should not be either too
large or too small. To make sure that the contributions from the
higher states are exponentially damped ( see Eq.
(\ref{eq:formre2})) and the global quark-hadron duality is
satisfied, we need a smaller Borel mass. On the other hand, the
Borel mass could not be too small for the  validity of OPE near
the light-cone for the correlation function, since the
contributions of higher twist distribution amplitudes amount to
the higher power of ${1 / M^2}$ to the perturbative part. In this
way, we  find a Borel platform $M^2\in [3.5,5]\rm{GeV}^2$. The
Borel mass dependence of the form factors is plotted  in
Fig.~\ref{fig:mtdependence} and Fig.~\ref{fig:higherstates}, the
former includes the contribution from the three-point $B$ meson
distribution amplitudes and the higher states contribution is
shown in the latter one. From these diagrams we can easily see
that the higher Fock state is highly suppressed in the Borel
window, and  higher exited states and the continuum states
contribution is within $15\%$ for $f_{D_0^*}^+(0)$($30\%$ for
$f_{D_0^*}^-(0)$). The numerical value for these form factors are
collected in Table \ref{tab:ffresults}, where the uncertainties
are from the combination of  the variation of shape parameter
$\omega_0$, the fluctuation of threshold value, the uncertainties
of quark masses and the errors of decay constants for the involved
mesons. The results in the other studies are listed for
comparison. We can see that for $f_{D_{s0}^*}^+(0)$ our result is
sightly larger than light-meson LQSR, however the results are
consistent with each other within the errors. For
$f_{D_{s0}^*}^-(0)$, the sign of our result is consistent with
that obtained from the QCDSR, but it is different from that
derived in the light meson LCSR. This discrepancy is expected to
be smeared by power corrections.

We can also investigate the $q^2$ dependence of the form factors
 $f_{D^{\ast}_{0}}(q^2)$.  It is known that
the OPE for the correlation function  is valid only at small
momentum transfer region
$0<q^2<(m_b-m_c)^2-2\Lambda_{\rm{QCD}}(m_b-m_c)$. At the large
momentum transfer region, we need to parameterize them in terms of
 phenomenological models. To achieve this goal we firstly
analyze the  form factors within the HQET framework, which works
well for the $b \to c $ transition. The matrix elements
responsible for $ B\to D^*_{0}$ transition can be parameterized as
\cite{Kurimoto:2002sb}
\begin{eqnarray}
 \langle D^{*+}_{0}(P)|\bar c\gamma_{\mu}\gamma_5 b|\bar B(P+q)\rangle
 &=&-i\sqrt{m_{B}m_{D^*_{0}}}[\eta_{D^*_{0}}^+(w)(v+v^{\prime})_{\mu}+\eta_{D^*_{0}}^-(w)(v-v^{\prime})_{\mu}],
 \label{eq:formHQET}
\end{eqnarray}
where $v=(P+q)/m_{B}$ and $v^{\prime}=P/m_{D^*_{0}}$ are the
four-velocity vectors of $B$ and $D^*_{0}$ mesons, and
 $w=v\cdot v^{\prime}=(m_{B}^2+m_{D^*_{0}}^2-q^2 )/
 2m_{B}m_{D^*_{0}}$.
Combining Eqs.~(\ref{formdef}) and (\ref{eq:formHQET}), we have
\begin{eqnarray}
 f_i^+(q^2)&=&\frac{1}{\sqrt{m_{B_i}m_{D_{i}}}}[(m_{B_i}+m_{D_{i}})\eta_i^+(w)-(m_{B_i}-m_{D_{i}})\eta_i^-(w)], \nonumber\\
 f_i^-(q^2)&=&\sqrt{\frac{m_{D_{i}}}{m_{B_i}}}[\eta_i^+(w)+\eta_i^-(w)],\label{eq:feta}
\end{eqnarray}
with $i= 1,2$ denotes strange and strangeless charmed scalar meson
respectively. Similarly to the Isgur-Wise function $\xi(v\cdot
v')$ for the $s$-wave transitions, heavy quark symmetry allows to
relate the form factors $\eta_{i}^{+}(w)$ and $\eta_{i}^{-}(w)$ to
a universal function $\tau_{1/2}(w)$\cite{Isgur}
\begin{eqnarray}
\eta_{i}^{+}(w)+\eta_{i}^{-}(w)=-2 \tau_{1/2}(w), \qquad
\eta_{i}^{+}(w)-\eta_{i}^{-}(w)=2 \tau_{1/2}(w).
\end{eqnarray}
Different  from  the Isgur-Wise function $\xi(w)$,  one can not
employ the  heavy quark symmetry to predict the normalization of
$\tau_{1/2}(w)$\cite{DeFazio:2000up}.

\begin{figure}
\begin{center}
\includegraphics[width=8.cm]{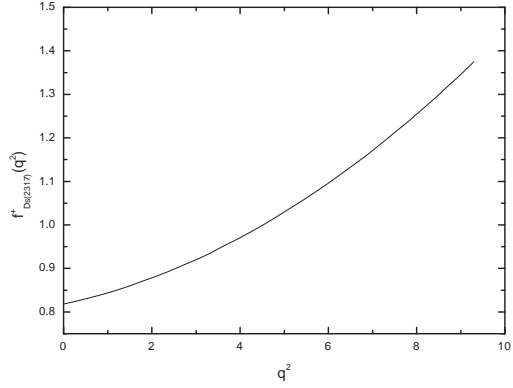}
\vspace{-0.5cm} \caption{ The dependence of form factor
$f_{D_s(2317)}^+$ on  $q^2$, }
 \label{q2dependence}
 \end{center}
 \end{figure}

Phenomenologically, one can parameterize the $B \to D^*_{0}$ form
factors in the small recoil region as
\begin{eqnarray}
\eta_i^{\pm}(w)&=&\eta_i^{\pm}(1)+a_i^{\pm}(w-1)+b_i^{\pm}(w-1)^2,\label{eq:paraHQET}
\end{eqnarray}
The parameters $\eta_i^{\pm}(1)$, $a_i^{\pm}$ and $b_i^{\pm}$ can
be determined by connecting the form factors derived in the LCSR
and HQET approaches  in the vicinity of region with $q^2 \sim
(m_b-m_c)^2-2\Lambda_{\rm{QCD}}(m_b-m_c)$. In this way, we can
derive the results of form factors in the whole kinematical
region, in Fig. (\ref{q2dependence}) we take $f_{D_{s0}}^+(q^2)$
as an example. The parameters related to all the form factors are
tabulated in Table \ref{tab:ffresults}.

As discussed before, the power-suppressed form factors $f^-_i$ in
Table \ref{tab:ffresults} suffer from sizable power corrections,
which can even change the sign.
 Generally speaking, the corrections can be picked up by perform the heavy
 quark
expansion of the current
\begin{eqnarray}
\bar{c} \Gamma_i b = \bar{c}_{v_2} \Gamma_i b_{v_1} - {1 \over 2
m_c} \bar{c}_{v_2} \Gamma_i i \not\! D_{\perp2} b_{v_1} + {1 \over
2 m_b} \bar{c}_{v_2} \Gamma_i i \not\!  D_{\perp1} b_{v_1} + ...
\end{eqnarray}
The last two terms in the above equation might give important
contribution for finite quark mass,  which  could help to  reduce
the discrepancy  among  different approaches. In addition, the
radiative correction may also help.

\begin{table}
\begin{center}
\caption{Numbers of $f_i^{\pm}(0)$ and $\eta_i^{\pm}(w)$
determined from the LCSR approach, where the uncertainties from
the Borel mass, threshold value, quark masses and decay constants
are combined together. For comparison,  the results estimated in
the QCDSR are also collected  here.} \label{tab:ffresults}
\begin{tabular}{cccc|rcccc}
\hline\hline
\ \ \                     & this work &Light meson LQSR    &QCDSR       $\;\;\;\;\;\;$    & &$\eta_i^{\pm}(1)$   &$a_i^{\pm}$       &$b_i^{\pm}$   \\
\hline
\ \ \  $f_{D^*_{s0}}^+(q^2)$ & $0.80_{-0.19}^{+0.24}$  &$0.53_{-0.11}^{+0.12}$   &$0.40 \pm 0.10$ \cite{Aliev:2006qy}      &$\eta_{D^*_{s0}}^+(w)$   &$0.29_{-0.06}^{+0.08}$      &$-0.49_{-0.54}^{+0.33}$      &$0.53_{-0.56}^{+0.86}$  \\
\ \ \  $f_{D^*_{s0}}^-(q^2)$ &$-0.20_{-0.10}^{+0.08}$     &$0.18_{-0.04}^{+0.06}$   &$-0.12 \pm 0.13$ \cite{Aliev:2006qy}     &$\eta_{D^*_{s0}}^-(w)$   &$-0.86_{-0.24}^{+0.23}$      &$1.59_{-0.42}^{+0.60}$      &$-1.61_{-0.90}^{+0.62}$  \\
\ \ \  $f_{D^*_0}^+(q^2)$  & $0.94_{-0.24}^{+0.31}$  &-  &-      &$\eta_{D^*_0}^+(w)$   &$0.28_{-0.07}^{+0.11}$      &$-0.34_{-0.65}^{+0.69}$      &$0.32_{-1.21}^{+1.10}$  \\
\ \ \  $f_{D^*_0}^-(q^2)$  &$-0.27_{-0.11}^{+0.12}$   &-   &-      &$\eta_{D^*_0}^-(w)$   &$-1.01_{-0.32}^{+0.26}$      &$1.86_{-0.62}^{+1.14}$      &$-2.00_{-1.72}^{+0.99}$  \\
\hline\hline
\end{tabular}
\end{center}
\end{table}

\section{Semileptonic decays }

The semileptonic  decays  $\bar  B_{(s)} \to D^*_{0(s)}l\nu$ are
important measurements in the $B$ factory which can be connected
with the form factors directly. The differential decay width is
given by:
 \begin{eqnarray}
 \frac{d\Gamma}{dq^2}&=&\frac{G_F^2|V_{cb}|^2}{768 \pi^3
 m_B^3}\frac{(q^2-m_l^2)^2}{(q^2)^3}\sqrt{\lambda}
 \bigg[\big(2m_l^2(\lambda+3q^2m_{D^*_0}^2)+q^2\lambda\big)|f_i^+(q^2)|^2\nonumber\\
 &&+6q^2m_l^2(m_B^2-m_{D^*_0}^2-q^2)f_i^+(q^2)f_i^-(q^2)+6q^4m_l^2|f_i^-(q^2)|^2\bigg],\label{eq:semil}
 \end{eqnarray}
with $\lambda=(m_B^2-m_{D^*_0}^2-q^2)^2-4q^2m_{D^*_0}^2$.

 \begin{figure}
 \begin{center}
 \includegraphics[width=6.5cm]{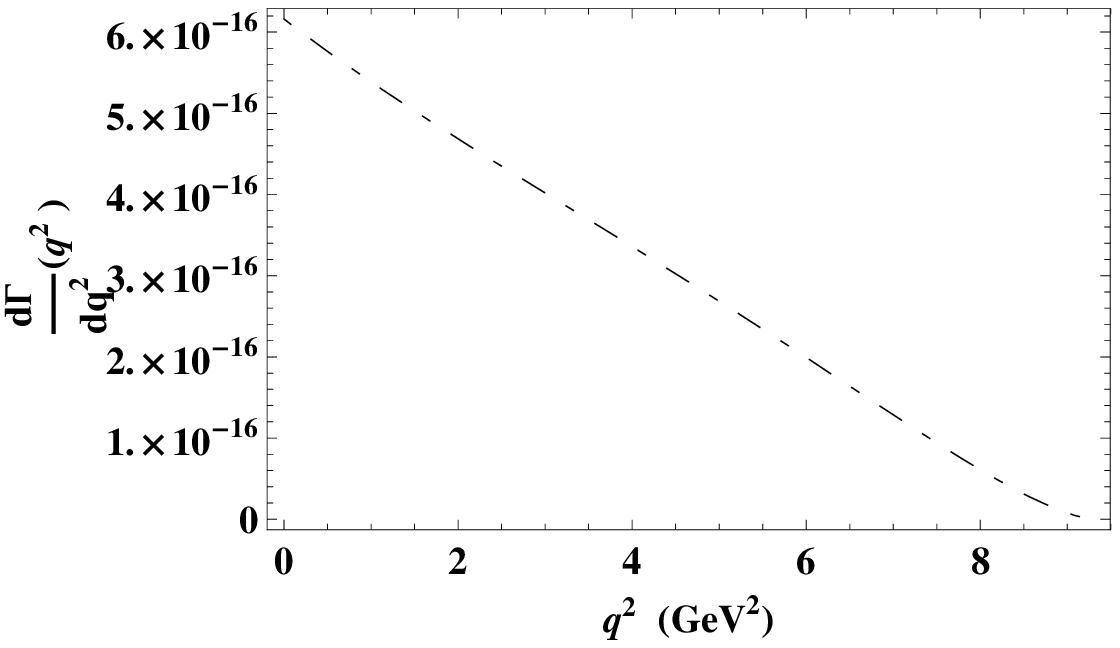}
 \hspace{1.0cm}
 \includegraphics[width=6.5cm]{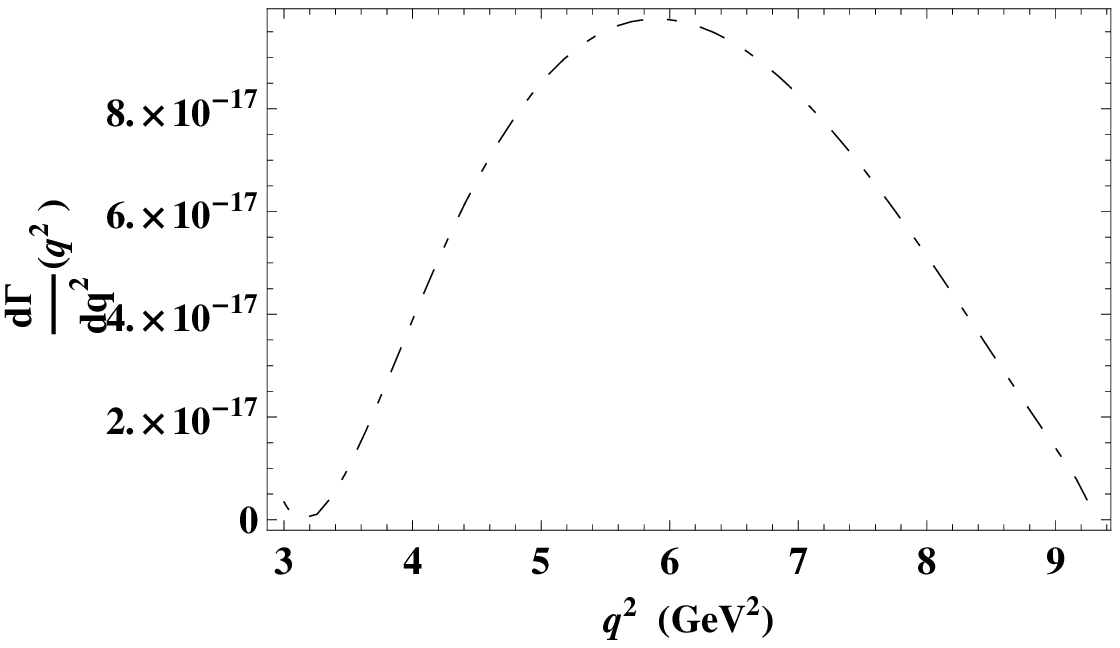}
 \vspace{-0.5cm}
\caption{The $q^2$ dependence of differential decay width
$\frac{d}{dq^2} \Gamma(\bar B_s^0\to D_{s0}^{*+}l^-\bar\nu_l)$ for
the final states with  $l=e,\mu$ (left figure)  and $l=\tau$
(right figure).}
 \label{fig:semiDs1}
 \end{center}
 \end{figure}

The $q^2$ dependence of these $\bar B_s^0\to
D_{s0}^{*+}l^-\bar\nu_l$ partial decay rates are plotted in Fig.
(\ref{fig:semiDs1}). Similar figures can also describe the $
B^-\to D_{0}^{*0}l^-\bar\nu_l$ decays. The curve
 of the $\tau$ final state is different from the light quark case for its mass effect.
 Integrating Eq. (\ref{eq:semil}), we get the branching fractions
of $\bar  B_{(s)} \to D^*_{0(s)}l\nu$ as grouped in Table
\ref{tab:Brforsemi}. The results from the constituent quark model,
the QCD sum rules and the light quark LCSR are also listed here.
Our result is slightly larger than the light quark LCSR as we have
obtained large form factors. Note that the theoretical error is
very large, which makes all the results are  actually consistent.
Besides, we can also find that the decay rates for the final state
with $\tau$ lepton are generally $3-4$ times smaller than those
for the muon case due to the suppression of phase spaces. The
branching fractions for $ \bar{B^0}\to D_0^{*0}(2400) l
\bar{\nu}_l$ are also available, which is the first prediction for
these decays, and we hope the future experiments can check our
results.
 \begin{table}
 \caption{Branching ratios for the semileptonic decays  $\bar B_s^0\to D_s^+(2317) l
\bar{\nu}_l$  and $ B^-\to D_0^{*0}(2400) l \bar{\nu}_l$ with the
form factors estimated in $B$-meson LCSR, where the results
calculated in the conventional light meson LCSR, the constituent
quark model and QCDSR are also displayed for comparison.}
 \label{tab:Brforsemi}
 \begin{center}
 \begin{tabular}{c c c}
 \hline\hline
 \ \ \ $\bar B_s^0\to D_{s0}^+l^-\bar\nu_l$         &$l=e,\mu$                             &$l=\tau$ \\
 \ \ \  this work                           &$(6.0\pm 1.9)\times10^{-3}$            &$(8.2_{-2.0}^{+1.8})\times10^{-4}$\\
  \ \ \  Light meson LCSR                           &$(2.3_{-1.0}^{+1.2})\times10^{-3}$            &$(5.7_{-2.3}^{+2.8})\times10^{-4}$\\
 \ \ \  QCDSR\cite{Aliev:2006qy}    &$\sim 10^{-3}$                                &$\sim 10^{-4}$  \\
 \ \ \  Constituent Quark Model\cite{Zhao:2006at}   &$(4.90-5.71)\times 10^{-3}$           &$$  \\
 \ \ \ QCDSR in HQET\cite{Huang:2004et}      &$(0.9-2.0)\times 10^{-3} $                                 &$$\\
 \hline\hline
 \ \ \  $\bar B^0\to D_0^{*0}l^-\bar\nu_l$        &$l=e,\mu$            &$l=\tau$ \\
 \ \ \  this work       &$(8.7_{-2.8}^{+5.1})\times10^{-3}$            &$(1.1^{+0.6}_{-0.3})\times10^{-3}$\\
  \hline\hline
 \end{tabular}
 \end{center}
 \end{table}

\section{Discussion and conclusion}

The charmed scalar meson spectroscopy has received  many research
interests both experimentally and theoretically. The $B_{(s)}$
decays provide  ideal places to study their property. In this
article, we employ the $B$-meson light-cone sum rules to compute
the $\bar B_s^0\to D_{s0}^{*+}(2317)$ and  $B^-\to D_0^{*0}(2400)$
transition form factors at large recoil region, assuming
$D_{s0}^{*+}(2317)$ and $D_0^{*0}(2400)$ being scalar
quark-anti-quark states. With the help of HQET, we extrapolate the
result to the whole momentum region, the $q^2$ dependence has beep
plotted in Fig(\ref{q2dependence}). Our results are compared with
the studies using the other nonperturbative methods, such as the
light-quark LCSR, the QCD sum rules and the quark models.
Considering large uncertainties, our results are consistent with
these studies. Meanwhile, we also found that the power corrections
should be large, which even change the sign of power-suppressed
form factor $f^-_{D_{s0}^{*+}(2317)}$.

Subsequently, we utilize the form factors obtained using $B$-meson
LCSR to estimate the semileptonic decays $\bar B_s^0\to
D_s^{*+}(2317)l \bar{\nu}_l$ and  $B^-\to D_0^{*0}(2400)l
\bar{\nu}_l$.  It has been shown in this work that the branching
fraction of the semileptonic $\bar B_s^0\to D_s^{*+}(2317)l
\bar{\nu}_l$ decay is around  $6\times 10^{-3} $ for light leptons
and   $0.8\times 10^{-3} $ for tau final state. The difference is
due to the phase space suppression. The predicted values can
confront with  the future LHCb measurements. The predicted
branching ration of $B^-\to D_0^{*0}(2400)l \bar{\nu}_l$ is
slightly larger than $\bar B_s^0\to D_s^{*+}(2317)l \bar{\nu}_l$,
and this observation can be  tested  at both LHCb and super $B$
factories.

\section*{Acknowledgement}
This work is partly supported by National Natural Science
Foundation of China under the Grant No. 11005100, 10735080 and
11075168. This research is also supported in part by the Project
of Knowledge Innovation Program(PKIP) of Chinese Academy of
Sciences, Grant No. KJCX2.YW.W10.

\section*{Appendix}
In the following we show the form factors from the 3-point B meson
DA.

\begin{eqnarray}
f^{3p}_\pm&=&\frac{f_{B}(m_c-m_q)}{2f_{D_{0}^*}m_{D^*_{0}}}
\{\int_0^{\eta_0m_B}d\omega\int_{\eta_0m_B-\omega}^{\infty}\frac{d\xi}{\xi}e^{-(s_0-m_{D_{0}^{*2}})/M_B^2}\nonumber\\
\nonumber &&\times
f(\eta_0)(m_{B}A^\pm_1+m_cA^\pm_2+A^\pm_3-A^\pm_4+A^\pm_5-\frac{A^\pm_6+A^\pm_7+m_BA^\pm_8}{M_B^2})\nonumber\\
&&+\int_0^{\eta_0}\frac{d\eta}{\bar{\eta}^2}\int_0^{\eta
m_B}d\omega\int_{\eta m_B-\omega}^{\infty}\frac{d\xi}{\xi}
\frac{1}{M_B^2}e^{-(s-m_{D_{0}^{*2}})/M_B^2}\nonumber\\
\nonumber &&\times(m_{B}B^\pm_1+m_cB^\pm_2+B^\pm_3-B^\pm_4+B^\pm_5-\frac{B^\pm_6+B^\pm_7+m_BB^\pm_8}{2M_B^2})\nonumber\\
 &&-\frac{f(\eta_0)e^{-(s_0-m_{D_{0}^{*2}})/M_B^{2}}}{2m_B^3}\int_0^{\eta_0m_B} d \omega\int_{\eta_0m_B-\omega}^{\infty}\frac{d\xi}{\xi}
 (C^\pm_4+C^\pm_6+C^\pm_7+C^\pm_8)\},
\end{eqnarray}
where the functions $A_i^\pm(i=1,2,...,8)$,$A_i^\pm(i=1,2,...,8)$
and $A_i^\pm(i=1,2,...,8)$ entering the integration are given
below:
\begin{eqnarray}
A^+_1&=&[\frac{2\alpha_0(2-3\eta_0+(s_0-q^2)/m_B^2)}{m_B\bar{\eta_0}^2}-\frac{1+2(s_0-q^2)/m_B^2+3(r_c-\eta_0)}{m_B\bar{\eta_0}^2}](\psi_A-\psi_V)
\nonumber \\
 A^+_2&=&\frac{6\alpha_0
\bar{\eta_0}-6r_c}{m_B^2\bar{\eta_0}^2}\psi_V\nonumber \\
A^+_3&=&\frac{-2\alpha_0 }{m_B\bar{\eta_0}^2}\bar{X_A}\nonumber\nonumber\\
A^+_4&=&\frac{\alpha_0(-m_B^2\bar{\eta}_0^2+m_c^2+q^2)}{m_B\bar{\eta}_0^3}\bar{X_A}\nonumber\\
A^+_5&=&\frac{\bar{\eta}_0+2r_c }{m_B\bar{\eta_0}^3}\bar{X_A}
\nonumber \\
A^+_6&=&\frac{2[m_B^2(\bar{\eta}_0-2r_c)+\frac{m_c^2-q^2}{\bar{\eta}_0}](\bar{\eta}_0+r_c)}{m_B\bar{\eta}_0^3}\bar{X_A}
\nonumber\\
A^+_7&=&\frac{-24\alpha_0 m_c^2}{m_B\bar{\eta}_0^3}\bar{X_A}
\nonumber \\
A^+_8&=&\frac{(-24m_c)[m_B^2(\bar{\eta}_0-2r_c)+\frac{m_c^2-q^2}{\bar{\eta}_0}](\bar{\eta}_0+r_c)}{m_B\bar{\eta}_0^3}\bar{Y_A}\nonumber
\\ A^-_1&=&[\frac{2\alpha_0(1-3\eta_0+(s_0-q^2)/m_B^2)}{m_B\bar{\eta_0}^2}-\frac{2+2(s_0-q^2)/m_B^2+3(r_c-\eta_0)}{m_B\bar{\eta_0}^2}](\psi_A-\psi_V)
\nonumber \\
 A^-_2&=&\frac{-6\alpha_0
{\eta_0}-6r_c}{m_B^2\bar{\eta_0}^2}\psi_V\nonumber \\
A^-_3&=&\frac{2\alpha_0(1+\eta_0) }{m_B\bar{\eta_0}^2}\bar{X_A}\nonumber\nonumber\\
A^-_4&=&\frac{\alpha_0(m_B^2\eta_0\bar{\eta}_0+\frac{1+\bar{\eta}_0}{\bar{\eta}_0}m_c^2-\frac{{\eta}_0}{\bar{\eta}_0}q^2)}{m_B\bar{\eta}_0^3}\bar{X_A}\nonumber\\
A^-_5&=&\frac{1+{\eta}_0-2r_c }{m_B\bar{\eta_0}^3}\bar{X_A}
\nonumber \\
A^-_6&=&\frac{2[m_B^2({\eta}_0-2r_c)+\frac{m_c^2-q^2}{\bar{\eta}_0}](-{\eta}_0+r_c)}{m_B\bar{\eta}_0^3}\bar{X_A}
\nonumber\\
A^-_7&=&\frac{-24\alpha_0 m_c^2}{m_B\bar{\eta}_0^3}\bar{X_A}
\nonumber \\
A^-_8&=&\frac{(-24m_c)[m_B^2({\eta}_0-2r_c)+\frac{m_c^2-q^2}{\bar{\eta}_0}](-{\eta}_0+r_c)}{m_B\bar{\eta}_0^3}\bar{Y_A}
\end{eqnarray}
\begin{eqnarray}
B_1^+&=&2\alpha(2-3\eta+(s-q^2)/m_B^2)-(1+2(s-q^2)/m_B^2+3(r_c-\eta))(\psi_A-\psi_V)
\nonumber \\
B_2^+&=&(6\alpha \bar{\eta}-6r_c)\psi_V \nonumber\\
B_3^+&=&(-2\alpha m_B )\bar{X_A} \nonumber\\
B_4^+&=&2\alpha m_B(-m_B^2\bar{\eta}^2+m_c^2+q^2)\bar{X_A}
\nonumber \\
B_5^+&=& (\bar{\eta}+2r_c) m_B )\bar{X_A}\nonumber \\
B_6^+&=&2[m_B^2(\bar{\eta}-2r_c)+\frac{m_c^2-q^2}{\bar{\eta}}](\bar{\eta}+r_c)\bar{X_A}
\nonumber \\
B_7^+&=&(-24\alpha m_Bm_c^2)\bar{X_A}
\nonumber \\
B_8^+&=&(-24m_c)[m_B^2(\bar{\eta}-2r_c)+\frac{m_c^2-q^2}{\bar{\eta}}](\bar{\eta}+r_c)\bar{Y_A}\end{eqnarray}
\begin{eqnarray}B_1^-&=&2\alpha(1-3\eta+(s-q^2)/m_B^2)-(2+2(s-q^2)/m_B^2+3(r_c-\eta))(\psi_A-\psi_V)
\nonumber \\
B_2^-&=&(-6\alpha {\eta}-6r_c)\psi_V \nonumber\\
B_3^-&=&\frac{2\alpha(1+\eta) }{m_B\bar{\eta}^2}\bar{X_A} \nonumber\\
B_4^-&=&2\alpha m_B
(m_B^2\eta\bar{\eta}+\frac{1+\bar{\eta}}{\bar{\eta}}m_c^2-\frac{\eta}{\bar{\eta}}q^2))\bar{X_A}
\nonumber \\
B_5^-&=& (1+{\eta}-2r_c) m_B )\bar{X_A}\nonumber \\
B_6^-&=&2[m_B^2(\bar{\eta}-2r_c)+\frac{m_c^2-q^2}{\bar{\eta}}](-{\eta}+r_c)\bar{X_A}
\nonumber \\
B_7^-&=&(-24\alpha m_Bm_c^2)\bar{X_A}
\nonumber \\
B_8^-&=&(-24m_c)[m_B^2(\bar{\eta}-2r_c)+\frac{m_c^2-q^2}{\bar{\eta}}](\bar{\eta}+r_c)\bar{Y_A}
\end{eqnarray}
\begin{eqnarray}
C^+_4&=&\frac{d}{d\eta}[2\alpha(-m_B^2\bar{\eta}^2+m_c^2+q^2)\frac{f(\eta)\bar{X_A}}{\bar{\eta}^3}]_{\eta=\eta_0}\nonumber\\
C^+_6&=&\frac{d}{d\eta}[2[m_B^2(\bar{\eta}-2r_c)+\frac{m_c^2-q^2}{\bar{\eta}}]
(\bar{\eta}+r_c)\frac{f(\eta)\bar{X_A}}{\bar{\eta}^3}]_{\eta=\eta_0}\nonumber\\
C^+_7&=&\frac{d}{d\eta}[(-24\alpha
m_c^2)\frac{f(\eta)\bar{Y_A}}{\bar{\eta}^3}]_{\eta=\eta_0}\nonumber\\
C^+_8&=&\frac{d}{d\eta}[(-24m_c)[m_B^2(\bar{\eta}-2r_c)+\frac{m_c^2-q^2}{-{\eta}}]
(\bar{\eta}+r_c)\frac{f(\eta)\bar{Y_A}}{\bar{\eta}^3}]_{\eta=\eta_0}\nonumber\\
C^-_4&=&\frac{d}{d\eta}[2\alpha(m_B^2\eta\bar{\eta}+\frac{1+\bar{\eta}}{\bar{\eta}}m_c^2-\frac{\eta}{\bar{\eta}}q^2)\frac{f(\eta)\bar{X_A}}{\bar{\eta}^3}]_{\eta=\eta_0}\nonumber\\
C^-_6&=&\frac{d}{d\eta}[2[m_B^2(\bar{\eta}-2r_c)+\frac{m_c^2-q^2}{\bar{\eta}}]
(-{\eta}+r_c)\frac{f(\eta)\bar{X_A}}{\bar{\eta}^3}]_{\eta=\eta_0}\nonumber\\
C^-_7&=&\frac{d}{d\eta}[(-24\alpha
m_c^2)\frac{f(\eta)\bar{Y_A}}{\bar{\eta}^3}]_{\eta=\eta_0}\nonumber\\
C^-_8&=&\frac{d}{d\eta}[(-24m_c)[m_B^2(\bar{\eta}-2r_c)+\frac{m_c^2-q^2}{-{\eta}}]
(-{\eta}+r_c)\frac{f(\eta)\bar{Y_A}}{\bar{\eta}^3}]_{\eta=\eta_0}\end{eqnarray}
Where the notations $\eta= \omega+\xi\alpha$, $ f(\eta)=
\left(1+\frac{m^2-q^2 }{\bar{\eta}^2 m_B^2}\right)^{-1}\,, $
$\overline{X}_A(\omega,\xi)=\int\limits_0^\omega d\tau
X_A(\tau,\xi),~~\overline{Y}_A(\eta,\xi)=\int\limits_0^\omega
d\tau Y_A(\tau,\xi),$ and $\eta_0$ satisfies the equation $
\bar{\eta}s_0-(\eta\bar{\eta}+r_c^2)m_{B}^2+\eta m_{B}^2=0$.


\end{document}